\documentclass[preprint,aip,rsi,amsmath,amssymb]{revtex4-1}
\usepackage{color}
\usepackage{amsmath}
\usepackage{graphicx}
\usepackage{multirow}
\usepackage{amsfonts}
\usepackage{amssymb}
\usepackage{amscd}

\begin{document}

\title{68 Gbps quantum random number generation by measuring laser phase fluctuations}

\author{You-Qi Nie}
\affiliation{Hefei National Laboratory for Physical Sciences at the Microscale and Department
of Modern Physics, University of Science and Technology of China, Hefei, Anhui 230026, China}
\affiliation{CAS Center for Excellence and Synergetic Innovation Center in Quantum Information
and Quantum Physics, University of Science and Technology of China, Hefei, Anhui 230026, China}

\author{Leilei Huang}
\affiliation{Department of Engineering Science, University of Oxford, Parks Road, Oxford, OX1 3PJ, UK}

\author{Yang Liu}
\affiliation{Hefei National Laboratory for Physical Sciences at the Microscale and Department
of Modern Physics, University of Science and Technology of China, Hefei, Anhui 230026, China}
\affiliation{CAS Center for Excellence and Synergetic Innovation Center in Quantum Information
and Quantum Physics, University of Science and Technology of China, Hefei, Anhui 230026, China}

\author{Frank Payne}
\affiliation{Department of Engineering Science, University of Oxford, Parks Road, Oxford, OX1 3PJ, UK}

\author{Jun Zhang}
\email{zhangjun@ustc.edu.cn}
\affiliation{Hefei National Laboratory for Physical Sciences at the Microscale and Department
of Modern Physics, University of Science and Technology of China, Hefei, Anhui 230026, China}
\affiliation{CAS Center for Excellence and Synergetic Innovation Center in Quantum Information
and Quantum Physics, University of Science and Technology of China, Hefei, Anhui 230026, China}

\author{Jian-Wei Pan}
\affiliation{Hefei National Laboratory for Physical Sciences at the Microscale and Department
of Modern Physics, University of Science and Technology of China, Hefei, Anhui 230026, China}
\affiliation{CAS Center for Excellence and Synergetic Innovation Center in Quantum Information
and Quantum Physics, University of Science and Technology of China, Hefei, Anhui 230026, China}

\begin{abstract}
The speed of a quantum random number generator is essential for practical applications, such as high-speed quantum key distribution systems. Here, we push the speed of a quantum random number generator to 68 Gbps by operating a laser around its threshold level. To achieve the rate, not only high-speed photodetector and high sampling rate are needed, but also a very stable interferometer is required. A practical interferometer with active feedback instead of common temperature control is developed to meet requirement of stability. Phase fluctuations of the laser are measured by the interferometer with a photodetector, and then digitalized to raw random numbers with a rate of 80 Gbps. The min-entropy of the raw data is evaluated by modeling the system and is used to quantify the quantum randomness of the raw data. The bias of the raw data caused by other signals, such as classical and detection noises, can be removed by Toeplitz-matrix hashing randomness extraction. The final random numbers can pass through the standard randomness tests. Our demonstration shows that high-speed quantum random number generators are ready for practical usage.
\end{abstract}
\maketitle

\section{Introduction}
Random numbers are widely used in many applications such as cryptography, scientific simulations, and lotteries. The basic characteristics of true random numbers include unpredictability, irreproducibility and unbiasedness. Most of true random number generators are based on physical systems, in which the quantum random number generator (QRNG) is an important approach with randomness coming from the indeterministic nature of quantum physics. Over the last two decades, various QRNG schemes have been proposed and implemented \cite{Rarity94,Stefanov00,Jennewein00,Ma05,Dynes08,Wayne09,Wayne10,Furst10,Qi10,Gabriel10,Symul11,Jofre11,Wahl11,Bustard11,Ren11,Xu12,Marandi12,Li13,Nie14,
Sanguinetti14}. For instance, a beam splitter scheme \cite{Rarity94, Stefanov00, Jennewein00} by measuring the path selection of single photons has been often used especially in the early stage of the development, and commercial QRNG products based on this scheme can be found in the market such as the IDQ Quantis that can generate a random bit rate of 4 Mbps. Time measurement schemes \cite{Ma05, Wayne09, Wayne10, Wahl11, Li13, Nie14} by measuring and digitizing the arrival time of single photons have been recently proposed. In such schemes, the random bit rate can in principle be improved by multiple times compared with the beam splitter scheme, depending on the time resolution. Experiments have demonstrated that the random bit rates using this scheme can reach round 100 Mbps \cite{Wayne10, Wahl11, Nie14}. However, in both of the above schemes, the speed is mainly limited by the count rates of single-photon detectors.

Bit rate is the key parameter to characterize the performance of QRNG. To significantly increase the speeds of QRNGs, new schemes have been recently
proposed including vacuum state fluctuations \cite{Gabriel10,Symul11,Jofre11} and quantum phase fluctuations \cite{Qi10,Xu12}. In the scheme of quantum phase fluctuations \cite{Qi10,Xu12}, conventional photodetectors rather than single-photon detectors are used to measure quantum effects. The photons coming from a laser originate from two mechanisms, i.e., stimulated emission and spontaneous emission. In the standard quantum optics model, photons from stimulated emission are normally assumed to have fixed phases, whereas photons from spontaneous emission have random phases. Thus, the total phase of photons emitted from the laser always fluctuates over time. When the laser is operated around its threshold level, the ratio between spontaneous and stimulated emissions could be high enough for measurement, which implies that quantum noise may dominate the phase fluctuations in this scenario. Once phase fluctuations are converted into intensity fluctuations, the quantum phase noise can be easily measured using classical photodetectors and hence quantum randomness is generated. Since photodetectors are much faster than single-photon detectors, bit rates of QRNGs by measuring quantum phase fluctuations could be considerably high. Recently, the Toronto group have experimentally demonstrated QRNGs with bit rates of 500 Mbps \cite{Qi10} and 6 Gbps \cite{Xu12} using the scheme of quantum phase fluctuations. However, these rates are still not high enough for some applications. For instance, given a 10 GHz clocked quantum key distribution (QKD) system \cite{Takesue06}, it requires roughly a 40 Gbps random bit stream. Hence, developing a high-speed and low-cost QRNG for such QKD system is highly required.

In this paper, we report a 68 Gbps QRNG based on the scheme of quantum phase fluctuations.
The key device in our QRNG system is an actively stabilized interferometer. One output port of the interferometer
is monitored by an optical power meter, which sends feedback signals in real-time to precisely tune a phase shifter in one arm of the interferometer.
In this way, the interferometer can be stable even without temperature control. The phase fluctuations of photons are converted into intensity fluctuations, which are then measured with a high-speed photodetector. The voltage amplitudes of the photodetector are digitalized as raw random data with a maximum sampling frequency of 10 GHz, which is higher than previous implementation with one order of magnitude \cite{Xu12}.
We then develop a model for our QRNG system to evaluate the min-entropy of the raw random data, and apply Toeplitz-matrix hashing for randomness extraction \cite{Ma13}.
After post-processing, all the random bit streams can well pass the standard randomness tests.

\section{QRNG scheme and experimental setup}
Phase fluctuations, or phase noise in the frequency domain, of photons emitted from a laser originate from spontaneous emissions and are random in nature, which can be used as a quantum random source. Most importantly, phase fluctuations are measurable via an interferometer by converting phase into intensity. However, quantum phase fluctuations are often mixed with classical noise. To make the readout easier, the ratio between quantum noise and classical noise should be as high as possible. In this experiment, when the laser is operated at the threshold level quantum noise will be dominant. After modeling and quantifying the contribution of quantum noise, quantum random bits are finally generated.

\begin{figure}
\centering
\includegraphics[width=12 cm]{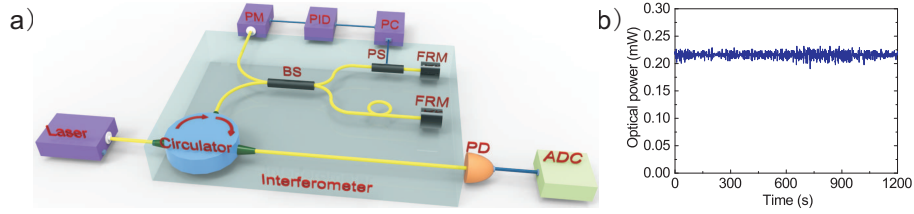}\\
\caption{a) Experimental setup. Photons emitted from a CW laser enter a polarization-insensitive Michelson interferometer after passing a circulator. One output port of the interferometer is detected by a PD after the circulator. The other output port is monitored by an optical PM. A PID program reads the measured values
of PM and sends feedback voltage signals to PC to adjust the PS. The measured amplitudes of PD are sampled by in build-in 8-bit ADC in a high-speed oscilloscope.
BS: beam splitter; PS: phase shifter; PM: power meter; PC: piezo controller; FRM: Faraday rotator mirror; PID: proportional-integral-derivative control; PD: photodetector; ADC: analog-to-digital converter. b) Measured power at OUT1 of the interferometer over 20 minutes, which indicates that the active phase stabilization for the interferometer works well.}
\label{fig1}
\end{figure}

The experimental setup is shown in Fig.~\ref{fig1}a.
A 1550 nm laser diode is driven by constant current with thermoelectric cooling control.
The continuous wavelength (CW) laser is operated slightly above its threshold.
Photons emitted from the laser enter a homemade Michelson interferometer.
The interferometer consists of a 3-port circulator, a 50/50 beam splitter (BS), a phase shifter (PS)
and two faraday rotator mirrors (FRMs). The time delay of about 3.735 ns
between the two arms of the interferometer is much smaller than the coherence time of laser.
With such a configuration, two FRMs can effectively remove polarization effects, which makes the interferometer
polarization-insensitive. The interferometer has two output ports, i.e., OUT1 and OUT2, as shown in Fig.~\ref{fig1}a.
OUT1 is monitored by an optical power meter (PM). The measured results of the PM are uploaded to a computer, in which a proportional-integral-derivative (PID) algorithm is implemented.
The PID program sends feedback signals to an open-loop piezo controller (PC, Thorlabs MDT694B) to precisely tune the PS in real-time. In such a way, phase stabilization in the interferometer could maintain at a high level and conventional temperature control for the interferometer is not required any more. Fig.~\ref{fig1}b shows a typical measurement result of optical power at OUT1 over 20 minutes. The nearly straight line in Fig.~\ref{fig1}b indicates that the interferometer itself is stable and slight fluctuations
could be regarded as classical noise.
OUT2 is detected by a 15 GHz photodetector (PD, EOT ET3500F).
The voltage output of the PD is sampled by an 8-bit analog-to-digital converter (ADC) in a 12 GHz oscilloscope (Agilent DSA91204A) and raw data are generated. Due to the memory limit in the oscilloscope, 20.5 Mpoints or 164 Mbits are acquired each time to form one raw data file. For a sampling rate of 10 GSa/s, the corresponding data acquisition period is 2.05 ms. The raw data is then transferred to a computer for post-processing.

\section{Min-entropy analysis and post-processing}
Here, we evaluate the randomness of the raw data from the QRNG following analysis in the literature \cite{Qi10,Xu12}. First, the electric field of the CW laser can be modeled as,
\begin{equation} \label{QRNG:Efield}
\begin{aligned}
E(t)=E_0\exp[i(\omega_0 t+\theta (t))].
\end{aligned}
\end{equation}
Then the interferometer output intensity (after interference) with a time delay of $\Delta T$ is
\begin{equation} \label{QRNG:I}
\begin{aligned}
2E^2_0+2E^2_0\cos\left[\omega_0\Delta T+\theta(t+\Delta T)-\theta (t)\right].
\end{aligned}
\end{equation}
Here, the phase fluctuations of the laser source are measured by the phase difference between time $t$ and $t+\Delta T$, defined as  $\Delta\theta(t)=\theta(t+\Delta T)-\theta (t)$.
After subtracting the direct current signal, the measured signal can be described by
\begin{equation} \label{QRNG:Icos}
\begin{aligned}
I(t)\propto P\cos(\omega_0\Delta T+\Delta\theta(t))=P[\cos(\omega_0\Delta T)\cos(\Delta\theta(t))-\sin(\omega_0\Delta T)\sin(\Delta\theta(t))].
\end{aligned}
\end{equation}
As shown in Fig.~\ref{fig1}a, the PS in the interferometer is constantly tuned such that $\omega_0\Delta T$ is $2m\pi+\pi/2$ ($m$ is an integer). Since $\Delta\theta(t)$ is small, we have
\begin{equation} \label{QRNG:Icossimple}
\begin{aligned}
I(t)\propto P\sin(\Delta\theta(t))\approx P\Delta\theta(t).
\end{aligned}
\end{equation}
Therefore, the phase fluctuation of the laser source can be measured directly by the intensity of the interferometer output. In addition to the signals from phase fluctuations, the variance of the photodetector output also contains background noise,
\begin{equation} \label{QRNG:v1}
\begin{aligned}
\langle V(t)^2\rangle = AP^2\langle \Delta\theta(t)^2\rangle+F,
\end{aligned}
\end{equation}
where $A$ is the linear response constant between the optical input signal and the photodetector voltage output. Meanwhile, as discussed previously the total phase fluctuations can be divided into signals from quantum fluctuations (Q/P) and classical ones (C) \cite{Qi10,Xu12},
\begin{equation} \label{QRNG:theta}
\begin{aligned}
\langle \Delta\theta(t)^2 \rangle =Q/P+C.
\end{aligned}
\end{equation}
Finally, there are three contributions to the measured intensity variance,
\begin{equation} \label{QRNG:v2}
\begin{aligned}
\langle V(t)^2\rangle = AQP+ACP^2+F.
\end{aligned}
\end{equation}
In practice, by changing the laser power $P$, the parameters of $AQ$, $AC$, and $F$ can be estimated via data fitting. Here, $AQP$ is the quantum signal of interest that should be maximized, compared with the classical counterpart $ACP^2+F$. With randomness extraction, one can remove the correlation between classical noise and the final random bits.

\begin{figure}
\centering
\includegraphics[width=8 cm]{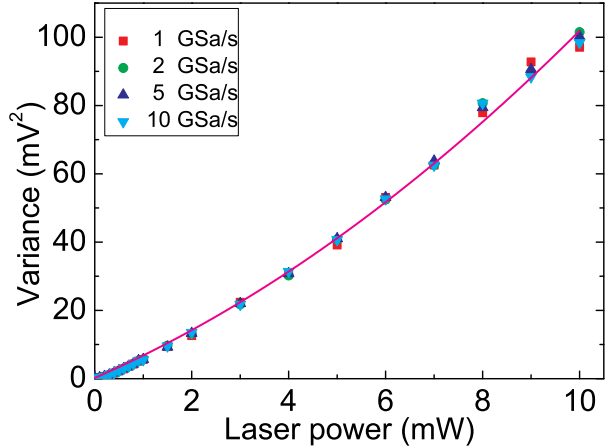}
\caption{The PD amplitude variance as a function of LD power with different sampling rate ranging from 1 GSa/s to 10 GSa/s. The solid line represents the theoretical fit with a sampling rate of 10 GSa/s.}
\label{fig2}
\end{figure}

The amplitude variances of the PD output are measured by precisely tuning the LD power from minimum to 10 mW. The PD output is sampled by a high-speed oscilloscope with different sampling rates. The time delay of the interferometer and the sampling time interval are two free variables that appear in the min-entropy evaluation model \cite{Ma15}. Taking into account that the bandwidths of PD and ADC are 15 GHz and 12 GHz, respectively, the maximum sampling rate in the experiment is chosen as 10 Gsa/s. The results are shown in Fig.~\ref{fig2}. The signal variance responses to the laser power with four different sampling rates are highly consistent, where the minor differences are caused by the access noise of the measurement, such as the bandwidth and resolution of the oscilloscope. The measured values are then fitted, according to Eq.~\eqref{QRNG:v2}, to obtain the parameters $AQ$, $AC$, and $F$. The results are listed in Table \ref{tab1}. As one can see, the fitting values match with each other in the four different sampling cases.

\begin{table}
\centering
\tabcolsep0.03in
\caption{\label{tab1} The theoretical fitting values of the parameters in the cases of four sampling rates.}
\begin{tabular}{c||c|c|c|c}
\hline
\hline
Parameter   &1 GSa/s    &2 GSa/s    &5 GSa/s    &10 GSa/s\\
\hline
AQ ($mV^2/mW$)      &6.0142 &5.8581 &6.0569 &6.2068\\
AC ($mV^2/mW^2$)    &0.4201 &0.4563 &0.4279 &0.3958\\
F  ($mV^2$)         &0.1714 &0.2052 &0.2200 &0.2162\\
\hline
$R^2$               &0.9977 &0.9988 &0.9989 &0.9982\\
\hline
\hline
\end{tabular}
\end{table}

In order to quantify the randomness of the raw data, a min-entropy evaluation model is developed based on our system. To maximize the min-entropy of the raw data, one can optimize the ratio of the quantum signals to other noises \cite{Xu12,Ma13}, which is defined as,
\begin{equation} \label{QRNG:gamma}
\begin{aligned}
\gamma=\frac{AQP}{ACP^2+F}.
\end{aligned}
\end{equation}
This allows the value of $\gamma$ to be calculated based on the fitted values of parameters listed in Table \ref{tab1} according to Eq.~\eqref{QRNG:gamma}. Meanwhile, to verify whether the fitted value of $\gamma$ is accurate, we employ an experimental approach \cite{Xu12} to directly measure this parameter. We take advantage of the fact that classical noise should dominate the phase fluctuation when the laser power is high enough. Therefore, one can assume that the quantum signal ratio, $\gamma$, is close to 0 when the LD is operated at its maximum power. Then the classical noise, $AC$, can be directly measured (upper bounded).
Note that any finite value of $\gamma$ in the high power regime would result in an underestimation of the min-entropy for evaluation, which is allowed since only the lower bound of the min-entropy is needed in data post-processing.

\begin{figure}
\centering
\includegraphics[width=8 cm]{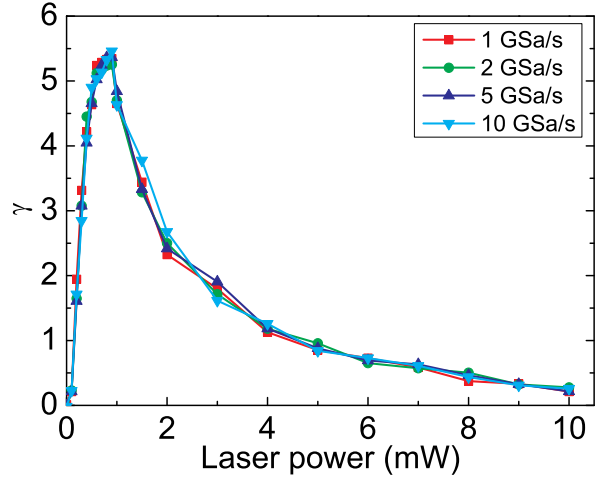}\\
\caption{The quantum signal ratio $\gamma$ as a function of LD power with different sampling rates. When the LD power is around 0.9 mW, $\gamma$ reaches the maximum value, 5.46.}
\label{fig3}
\end{figure}

In our experiment, the LD is operated at its maximum power (around 12 mW) and a digital variable attenuator is used before the interferometer to tune the incident power from minimum to 12 mW. Note that the key difference between this setup and the one to acquire data for Fig.~\ref{fig2} lies on the position of the variable attenuator. With this setup, the voltage variance due to the classical noise at each power can be directly measured. Combining the total voltage variance as measured in Fig.~\ref{fig2}, the experimental values of $\gamma$ in the cases of different laser powers and different sampling rates are obtained and shown in Fig.~\ref{fig3}. The experimental results agree with the theoretically calculated values. For our QRNG, the quantum signal ratio reaches its peak value, $\gamma=5.46$, at the power of 0.9 mW. In this optimal condition, signals due to quantum phase fluctuations dominate the sampling output. In the experiment, the LD is always operated at the optimal power to generate raw random data.

Combining Eq.~\eqref{QRNG:v2} and Eq.~\eqref{QRNG:gamma}, the variance of quantum signal is given by,
\begin{equation} \label{QRNG:qsigma}
\begin{aligned}
\sigma_{quantum}^2=AQP=\frac{\gamma}{\gamma+1}\langle V(t)^2\rangle.
\end{aligned}
\end{equation}
According to the theoretical model \cite{Ma13}, the quantum signal follows a Gaussian distribution. Thus, with its variance given in Eq.~\eqref{QRNG:qsigma}, one can obtain the whole distribution of the quantum signal. The randomness is quantified with min-entropy, defined as follows,
\begin{equation} \label{QRNG:Hmin}
\begin{aligned}
H_\infty(X)=-\log_2(\max_{x\in\{0, 1\}^n} P_r[X=x]).
\end{aligned}
\end{equation}
Then, the min-entropy of the raw data can be evaluated via the Gaussian distribution.

\begin{figure}
\includegraphics[width=13 cm]{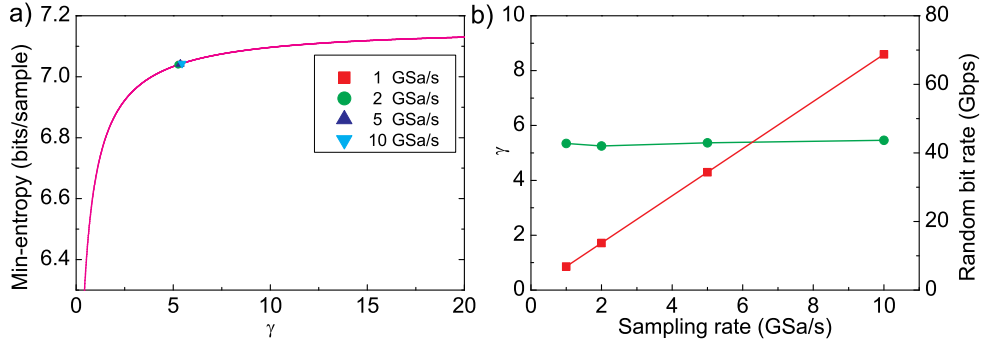}\\
\caption{a) The plot of min-entropy versus $\gamma$. Symbols and line represent experimental values and theoretical calculation, respectively.
b) Optimal values of $\gamma$ (left $y$-axis) and corresponding final random bit rates (right $y$-axis) in the cases of four sampling rates. With a sampling rate of 10 GSa/s, the random bit rate reaches 68 Gbps.}
  \label{fig4}
\end{figure}

In the experiment, the raw sampling data acquired in the oscilloscope are transferred to a computer for
further post-processing offline. A MATLAB program in the computer reads the data of 10 Mpoints each time. The program first sorts all the data and
finds the minimum and the maximum values to form the voltage range of $[V_{min}, V_{max}]$. Both the data in the ranges of the top $0.1\%$ and bottom $0.1\%$ are manually set to the corresponding boundaries so that the effective range is changed to $[V_{1}, V_{2}]$. This effective range is further digitized with 8 bits, corresponding to 256 bins, and all the remaining points are sorted out in these bins. In this way, the probability distribution of the raw data is formed and the bin that has the maximum probability can be easily found. The min-entropy is related to $\gamma$. The theoretical relationship between these two parameters is calculated according to the Gaussian distribution, as shown by the solid line of Fig.~\ref{fig4}a. The increase slope of min-entropy is steep when $\gamma$ is small, but becomes rather flat when $\gamma > 5$.
In the cases of four sampling rates, the values of $\gamma$ are measured as shown in the left y-axis of Fig.~\ref{fig4}b, and the values of min-entropy are obtained as well.

In the case of 0.9 mW laser power and 10 GSa/s sampling rate, the total voltage variance is $\langle V(t)^2\rangle=5.24~mV^2$ while the quantum part is $\sigma_{quantum}^2=4.49~mV^2$.
The effective voltage range is 10.2 mV and the maximum probability is $P_{max}=7.6\times10^{-3}$.
According to Eq.~\eqref{QRNG:Hmin}, the min-entropy in the case is 7.04 bits per sample point or $0.88$ bits per bit, which means that
7.04 random bits can be generated from each sample. Multiplying by the sampling rate of 10 Gsa/s, the resulting random bit rate is 70.4 Gbps.

A Toeplitz-matrix hash function based on fast Fourier transform (FFT) ~\cite{Ma13} with
a size of $n\times m$ ($n=3.36\times 10^7$ and $m= 2.88\times 10^7$) is applied, which means that $m$ final random bits are generated from $n$ raw random bits.
Therefore, the min-entropy is slightly decreased to 0.86 bits per bit ($m/n$) so that the final bit rate reaches 68 Gbps.
Compared with previous realizations based on quantum phase fluctuation schemes \cite{Qi10,Xu12} and other QRNG implementations
\cite{Rarity94, Stefanov00, Jennewein00, Ma05, Wayne09, Wayne10, Wahl11, Li13, Nie14}, the bit rate of our QRNG has been drastically increased.

Since the values of quantum-to-noise ratio $\gamma$ are almost constant for different sampling rates, the final random bit rate is linearly proportional to the sampling rate as shown in the right $y$-axis of Fig.~\ref{fig4}b. Here, we want to point out that the real-time QRNG speed might be further limited by the implementation of data post-processing --- randomness extraction. In our post-processing, the Toeplitz-matrix hash function is implemented by a MATLAB program on a personal computer \cite{Sup1}. The testing real-time randomness extraction rate of the program is about 1.6 Mbps. If the extraction is implemented by hardware, such as fast field-programmable gate arrays (FPGAs), the rate might reach a few hundreds of Mbps, or even Gbps regime. With multiple randomness extraction processors in parallel, a post-processing implementation that can match our QRNG is likely to be feasible. This is a very interesting perspective research topic for electric engineering.

\begin{figure}
\includegraphics[width=13 cm]{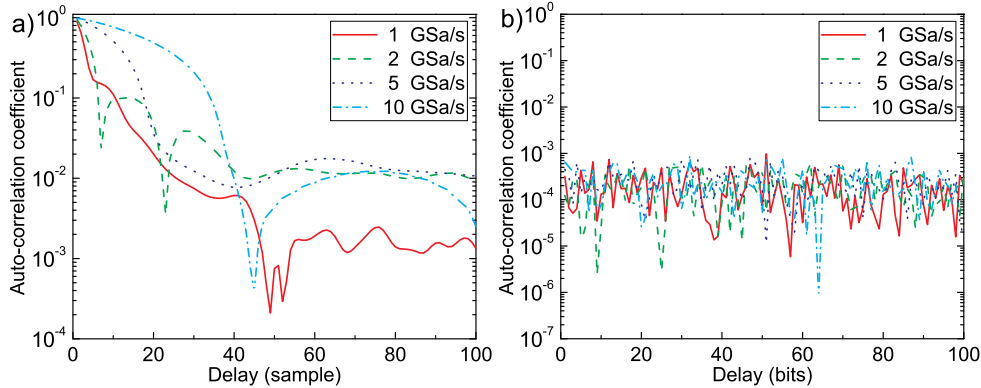}\\
\caption{a) The autocorrelations of the raw data with different sampling rates. Data size of each file is $10^7$ samples and each sample consists of 8 bits. b) The autocorrelations of the Toeplitz-hashing output. Data size of each file is $10^7$ bits. The autocorrelations are substantially reduced by Toeplitz-hashing.}
  \label{fig5}
\end{figure}

Classical noise can introduce bias and autocorrelation to the raw data. Thus, the raw data fails to pass half of the items in the NIST statistical tests~\cite{NIST}. The autocorrelation results of the raw data and final random bits after Toeplitz-hashing are shown in Fig.~\ref{fig5}. There are no significant differences between different sampling rates ranging from 1 GSa/s to 10 GSa/s ~\cite{Sup2}. The relatively large autocorrelation coefficient of the raw data as shown in Fig.~\ref{fig5}a is due to the fact that sampling time interval is shorter than the time delay between two arms of the interferometer in the experiment. This result is consistent with the theoretical model of the QRNG \cite{Ma15}. After Toeplitz-hashing, the autocorrelation is substantially reduced as shown in Fig.~\ref{fig5}b, and it is consistent with the scope of statistical fluctuations. It indicates that Toeplitz-hashing extraction is effective in the data post-processing for our experiment.

To further test and verify the quality of the final output random bits, we apply the standard NIST statistical tests~\cite{NIST}. The $p$-values and the proportions of all the test items in four cases are listed in Table~\ref{tab2}. For the test items that produce multiple outcomes, a Kolmogorov-Smirnov (KS) uniformity test is further performed for the $p$-values, and the proportions are averaged. Clearly, all the data passes the NIST tests.

\begin{table}
\centering
\tabcolsep0.01in
\caption{\label{tab2}. The results of the standard NIST tests for final random bits. Data size of each bits string is 1 Gbits. In the tests producing multiple outcomes, a Kolmogorov-Smirnov (KS) uniformity test has been further performed for $p$-values, and the corresponding proportions are averaged. All the $p$-values are larger than 0.01 and all the proportions are larger than 0.98, which indicates that random bits well pass the NIST tests.}
\begin{tabular}{l|c|c|c|c|c|c|c|c}
\hline
\hline
\multirow{2}{*}{Statistical test} &\multicolumn{2}{|c}{1 GSa/s}&\multicolumn{2}{|c}{2 GSa/s}&\multicolumn{2}{|c}{5 GSa/s}&\multicolumn{2}{|c}{10 GSa/s}\\
\cline{2-9}
                                  &P-value &Proportion    &P-value &Proportion    &P-value &Proportion    &P-value &Proportion\\
\hline
Frequency                         &0.4731  &0.991         &0.6288  &0.994         &0.1209  &0.987         &0.6288  &0.990\\
Block Frequency                   &0.5728  &0.985         &0.4172  &0.991         &0.2940  &0.991         &0.2925  &0.992\\
Cumulative Sum                    &0.5667  &0.988         &0.3770  &0.993         &0.6662  &0.985         &0.4313  &0.991\\
Runs                              &0.8360  &0.986         &0.9915  &0.992         &0.9932  &0.992         &0.6121  &0.993\\
Longest Run                       &0.2983  &0.992         &0.5362  &0.993         &0.0401  &0.986         &0.0904  &0.991\\
Rank                              &0.3299  &0.992         &0.1495  &0.993         &0.7792  &0.990         &0.2111  &0.992\\
FFT                               &0.9335  &0.983         &0.0259  &0.988         &0.3056  &0.989         &0.5281  &0.992\\
Non Overlapping Template          &0.0662  &0.990         &0.0507  &0.990         &0.1331  &0.990         &0.0599  &0.990\\
Overlapping Template              &0.4065  &0.991         &0.6080  &0.987         &0.8963  &0.987         &0.0707  &0.984\\
Universal                         &0.0366  &0.992         &0.1917  &0.990         &0.4885  &0.987         &0.7157  &0.986\\
Approximate Entropy               &0.0721  &0.986         &0.2417  &0.988         &0.0469  &0.991         &0.5141  &0.994\\
Random Excursions                 &0.3666  &0.991         &0.2875  &0.989         &0.1214  &0.989         &0.3026  &0.991\\
Random Excursions Variant         &0.1849  &0.991         &0.1810  &0.991         &0.2557  &0.987         &0.1407  &0.991\\
Serial                            &0.7439  &0.994         &0.6579  &0.989         &0.5997  &0.990         &0.6559  &0.992\\
Linear Complexity                 &0.9229  &0.992         &0.3086  &0.993         &0.1816  &0.990         &0.5976  &0.991\\
\hline
\hline
\end{tabular}
\end{table}

\section{Conclusion}
In conclusion, we have demonstrated a 68 Gbps QRNG based on quantum phase fluctuations. With the help of an actively stabilized interferometer, the laser phase fluctuations can be converted into intensity fluctuations for sampling and further digitizing. The QRNG system works stably without temperature control. We experimentally measured the ratio of quantum noise to classical noise so that the min-entropy of the raw data can be effectively evaluated. Toeplitz-matrix hashing is then used to extract final random bits, which pass the standard NIST tests well. With a sampling rate of 10 GSa/s, the final random bit rate reaches 68 Gbps, which shows a dramatic improvement compared with existing QRNG implementations. Our QRNG could be a practical approach for some specific applications such as QKD systems with a clock rate of over 10 GHz. An interesting question to ask is what is the limitation of this QRNG, given high enough precision of the data acquisition system and interferometer. Also, in our scheme the LD can be replaced with light-emitting diode (LED).

\begin{acknowledgments}
We acknowledge helpful discussions with J.-Y. Guan, W.-H.~Jiang, H.~Liang, C.~Wu, F.~Xu, Q.~Zhang, and X.-G. Zhang, especially X.~Ma for his enlightening inputs on the theoretical aspect of the subject. This work has been financially supported by the National Basic Research Program of China Grant No.~2013CB336800, the National High-Tech R\&D Program Grant No.~2011AA010802, the National Natural Science Foundation of China Grant No.~61275121, and the Chinese Academy of Sciences. Y.-Q.~Nie and L.~Huang contribute equally to this work.
\end{acknowledgments}


\begin{thebibliography}{99}

\bibitem{Rarity94}
J. Rarity, P. Owens, and P. Tapster,
J. Mod. Opt. \textbf{41}, 2435 (1994).

\bibitem{Stefanov00}
A. Stefanov, N. Gisin, O. Guinnarda, L. Guinnarda, and H. Zbinden,
J. Mod. Opt. \textbf{47}, 595 (2000).

\bibitem{Jennewein00}
T. Jennewein, U. Achleitner, G. Weihs, H. Weinfurter, and A. Zeilinger,
Rev. Sci. Instrum. \textbf{71}, 1675 (2000).

\bibitem{Ma05}
H.-Q. Ma, Y. Xie, and L.-A. Wu,
Appl. Optics \textbf{44}, 7760 (2005).

\bibitem{Dynes08}
J. Dynes, Z. Yuan, A. Sharpe, and A. Shields,
Appl. Phys. Lett. \textbf{93}, 031109 (2008).

\bibitem{Wayne09}
M. A. Wayne, E. R. Jeffrey, G. M. Akselrod, and P. G. Kwiat,
J. Mod. Opt. \textbf{56}, 516 (2009).

\bibitem{Wayne10}
M. A. Wayne and P. G. Kwiat,
Opt. Express \textbf{18}, 9351 (2010).

\bibitem{Furst10}
M. F\"urst, H. Weier, S. Nauerth, D. Marangon, C. Kurtsiefer, and H. Weinfurter,
Opt. Express \textbf{18}, 13029 (2010).

\bibitem{Qi10}
B. Qi, Y.-M. Chi, H.-K. Lo, and L. Qian,
Opt. Lett. \textbf{35}, 312 (2010).

\bibitem{Gabriel10}
C. Gabriel, C. Wittmann, D. Sych, R. Dong, W. Mauerer, U. Andersen, C. Marquardt, and G. Leuchs,
Nat. Photon. \textbf{4}, 711 (2010).

\bibitem{Symul11}
T. Symul, S. Assad, and P. Lam,
Appl. Phys. Lett. \textbf{98}, 231103 (2011).

\bibitem{Jofre11}
M. Jofre, M. Curty, F. Steinlechner, G. Anzolin, J. P. Torres, M. W. Mitchell, and V. Pruneri,
Opt. Express \textbf{19}, 20665 (2011).

\bibitem{Wahl11}
M. Wahl, M. Leifgen, M. Berlin, T. Rohlicke, H.-J. Rahn, and O. Benson,
Appl. Phys. Lett. \textbf{98}, 171105 (2011).

\bibitem{Bustard11}
P. J. Bustard, D. Moffatt, R. Lausten, G.Wu, I. A. Walmsley, and B. J. Sussman,
Opt. Express \textbf{19}, 25173 (2011).

\bibitem{Ren11}
M. Ren, E. Wu, Y. Liang, Y. Jian, G. Wu, and H. Zeng,
Phys. Rev. A \textbf{83}, 023820 (2011).

\bibitem{Xu12}
F. Xu, B. Qi, X. Ma, H. Xu, H. Zheng, and H.-K. Lo,
Opt. Express \textbf{20}, 12366 (2012).

\bibitem{Marandi12}
A. Marandi, N. C. Leindecker, K. L. Vodopyanov, and R. L. Byer,
Opt. Express \textbf{20}, 19322 (2012).

\bibitem{Li13}
S. Li, L. Wang, L.-A. Wu, H.-Q. Ma, and G.-J. Zhai,
J. Opt. Soc. Am. A \textbf{30}, 124 (2013).

\bibitem{Nie14}
Y.-Q. Nie, H.-F. Zhang, Z. Zhang, J. Wang, X. Ma, J. Zhang, and J.-W. Pan,
Appl. Phys. Lett. \textbf{104}, 051110 (2014).

\bibitem{Sanguinetti14}
B. Sanguinetti, A. Martin, H. Zbinden, and N. Gisin,
Phys. Rev. X \textbf{4}, 031056 (2014).

\bibitem{Takesue06}
H. Takesue, E. Diamanti, C. Langrock, M. M. Fejer, and Y. Yamamoto,
Opt. Express \textbf{14}, 9522 (2006).

\bibitem{Ma13}
X. Ma, F. Xu, H. Xu, X. Tan, B. Qi, and H.-K. Lo,
Phys. Rev. A \textbf{87}, 062327 (2013).

\bibitem{Ma15}
H. Zhou, X. Yuan, and X. Ma,
\emph{Submitted} (2015).

\bibitem{Sup1}
See Supplemental Material at [URL will be inserted by AIP] for the implementation of Toeplitz-matrix hash function based on fast Fourier transform.

\bibitem{Sup2}
Samples of raw random data files and the corresponding generated random files after post-processing with different sampling rates are uploaded on a public server, see
Supplemental Material at [URL will be inserted by AIP] for details.

\bibitem{NIST}
A. Rukhin, J. Soto, J. Nechvatal, M. Smid, E. Barker, S. Leigh, M. Levenson, M. Vangel,
D. Banks, A. Heckert, J. Dray, and S. Vo, NIST, Special Publication 14, Revision 1a (2010).

\end{thebibliography}
\end{document}